\documentclass[12pt]{article}
 
\usepackage{times}
\usepackage{geometry}
\usepackage[utf8]{inputenc}
\usepackage{enumitem,amssymb}
\usepackage{ragged2e}
\newlist{thematic}{itemize}{8}
\setlist[thematic]{label=$\square$}
\usepackage{pifont}

\usepackage{natbib}
\usepackage{aastex_hack}
\usepackage{graphicx}

\newcommand{\hmol}{H$_2$}
\newcommand{\1}{~{\sc i}}   
\newcommand{\2}{~{\sc ii}}
\newcommand{\3}{~{\sc iii}}
\newcommand{\4}{~{\sc iv}}
\newcommand{\5}{~{\sc v}}
\newcommand{\6}{~{\sc vi}}
\newcommand{\kms}{km~s$^{-1}$}

\begin{document}
\raggedright
\huge
Astro2020 Science White Paper \linebreak

Far- to near-UV spectroscopy of the interstellar medium  at very high resolution and very high signal-to-noise ratio\linebreak
\normalsize

\noindent \textbf{Thematic Areas:} \hspace*{57pt} 
$\square$ Planetary Systems \hspace*{13pt} 
$\square$ Star and Planet Formation \hspace*{20pt}\linebreak
$\square$ Formation and Evolution of Compact Objects \hspace*{31pt} 
$\square$ Cosmology and Fundamental Physics \linebreak
$\square$ Stars and Stellar Evolution \hspace*{1pt} 
$\boxtimes$ Resolved Stellar Populations and their Environments \hspace*{40pt} \linebreak
$\boxtimes$ Galaxy Evolution   \hspace*{45pt}
$\square$ Multi-Messenger Astronomy and Astrophysics \hspace*{65pt} \linebreak
  
\textbf{Principal Author:}

Name: Cecile Gry
 \linebreak						
Institution: LAM,  Aix Marseille Univ, CNRS, CNES,  Marseille, France
 \linebreak
Email: cecile.gry@lam.fr
 \linebreak
Phone:   +33 4 91 05 59 21
 \linebreak
 
\textbf{Co-authors:} \\
Edward B. Jenkins (Princeton University Observatory)\linebreak 
Patrick Boiss\'e (Institut d'Astrophysique de Paris) \linebreak
Jacques Le Bourlot (LERMA, Observatoire de Paris and Universite Paris-Diderot, Paris 7)\linebreak
Vianney Lebouteiller (Laboratoire AIM, CEA, Universite Paris-Saclay)\linebreak
Daniel Welty (Space Telescope Science Institute, Baltimore) \linebreak
Fran\c{c}ois Boulanger  (Laboratoire de Physique, Ecole Normale Sup\'erieure,  Paris)
  \linebreak

\textbf{Abstract :}
A comprehensive study of interstellar medium (ISM) phases and the nature of their boundaries or connections requires comparing abundances and velocity profiles from tracers of the different phases. In the UV, studies of a wealth of absorption features appearing in the spectra of hot stars yield fundamental insights into the composition and physical characteristics of all phases of the ISM along with the processes that influence them. They also inform us on the nature of boundaries between them. However no single instrument has as yet given access to species in all ISM phases at the same high spectral resolution: from the molecular bands of CO and \hmol\ in the far-UV, to the cold neutral medium tracers C\1\ and S\1\ and the warm medium tracers like C\2, N\1, O\1, Mg\2, Fe\2, Si\2\ etc..., and to the high ions of the hot ionized medium C\4, Si\4\ in the UV, as well as O\6\ in the far UV. We have yet to design the spectrometer that will enable observing the full UV domain at resolving power R$>$200\,000 and signal-to-noise ratio SNR$>$500. The line FWHM being governed by turbulence, temperature, and species mass, such a resolution is necessary to resolve lines from both the cold molecular hydrogen and the warm metal ions with a turbulence of $\sim 1$ \kms, and to differentiate distinct velocity components, typically separated by less than 2 \kms. 

\pagebreak
\section{Introduction: The multi-phase ISM of the Milky Way} \label{sec:intro}
Matter in the interstellar space is distributed in diverse, but well defined phases that consist of (1) the hot, ionized (T$\sim$10$^{6-7}$K) interstellar medium (HIM) emitting soft X-rays, (2) the warm neutral or ionized medium (WNM+WIM) (6000$<$T$<$10$^4$\,K), and (3) the cold (T$\sim$10-200K) neutral medium (CNM) and molecular star-forming clouds that occupy only a few percent of the volume. The different phases are supposed to be globally consistent with hydrostatic equilibrium \citep{Ferriere1998} however the multi-phase aspect of the ISM is not fully comprehended. How the different phases are related, and how mass flows from one phase to the other --in particular from H to \hmol\ which is the first step in stellar formation -- are still open questions. 
\section{Cold, dense and molecular gas}
 Observing lines in absorption is the only way to measure atoms and molecules in their ground state, which is particularly important for \hmol, found essentially in the J=0 and 1 levels in the ISM. Infrared vibrational emission lines concern a negligible fraction of the total \hmol.
  Therefore a direct estimate of the  mass of the cold molecular gas can be obtained only by observing \hmol\ in absorption in the far UV. 
  \linebreak
 Basic processes of interest in the molecular ISM include: (1) H-to-\hmol\ transition with a large sample of high-extinction targets, including the role of internal spatial structures for the self-shielding process of diffuse \hmol, (2) chemistry in truly molecular regions (molecular fraction f $\approx 1$), shielded from UV radiation by dust and \hmol\, where ionization by penetrating cosmic rays plays an important role, (3) change of molecular composition with $A_V$ by reaching opacities where HCN and HNC exist, (4) variation of the CO/\hmol\ ratio with cloud depth,  (5) characterization of the C-to-CO transition (occurring around $A_V=2$ depending on the ISRF)  with $A_V=5$ lines of sight , (6) internal velocity structure and measure of the turbulence in molecular clouds.
\linebreak
High resolution of the order of 1-2 km/s is required to differentiate the different components and separate fine-structure lines or rotational and vibrational levels, yet \hmol\ has been observed at a resolution better than 10 \kms\ only in a few lightly-reddened (thus low N(H2)) Galactic lines of sight \citep[with IMAPS, e.g.,][]{Jenkins.Peimbert1997}.  Paradoxically, higher resolution on \hmol\ is obtained more readily in the distant universe, due to redshifts that bring \hmol\ lines into the visible spectrum \citep[e.g.,][]{Noterdaeme2007}. High resolution is necessary in the far-UV to study the physical conditions, the excitation and formation mechanism of \hmol\ in the local universe. 
In more reddened sight lines it will be hard to distinguish individual
  components in the 
  J=0,1 lines which will be damped, but they could be resolved in the higher-J level lines.
\linebreak High sensitivity is also required to reach  high-extinction targets since 
we need to measure abundances  in clouds of different opacity, and access regions shielded from UV radiation by dust and \hmol\ where molecular composition changes dramatically and where the chemistry is influenced by penetration of cosmic rays. 
Extinction is very severe in the UV. Following \cite{Jenkins.Wallerstein1999}, for the usual gas-to-dust ratio, the log of the stellar flux at 1150\AA\  decreases by $\sim -$6.4 10$^{-22}\times$ N(H$_{tot}$), thus by $\sim -$1.2 A$_V$ relative to a non-reddened star. An extinction of A$_V$\,=\,4 produces an obscuration by a factor 60 000. Still, with an effective area three times that of COS, 20 minutes exposure time could provide a signal-to-noise ratio of 100 for stars comparable to the bright $\it Copernicus$ targets but obscured by this amount of material.
 \linebreak
Understanding of the basic processes at work in the cold gas also involves observing C\1, C\2, CO, and other molecules like CH and CH$^+$, relating them to \hmol\ \citep[e.g.,][]{Nehme2008} and comparing their column density and abundance ratios with detailed models (e.g., Meudon PDR model; \citealt{LePetit2006}). 
Carbon chemistry, in particular the abundance of C$^0$ can also be examined by looking for an expected discontinuity  in obscured spectra at 1101\AA\ due to carbon ionization \citep{Rollins.Rawlings2012}. For A$_V$=2 (resp. A$_V$=3) a flux decrease by a factor of 10 (resp. 100) is expected. Such a discontinuity depends on the relative abundance of C$^0$, governed by n$_H$, A$_V$ and chemistry. It has not been detected in FUSE spectra, and reaching higher A$_V$ is required to draw interesting conclusions.
%
\paragraph{Calibrating the CO/N(\hmol) relation}
The CO emission is widely used to trace \hmol\ but the CO-to-\hmol\ conversion factor X$_{\rm CO}$ is known to depend on metallicity  and most of the time it relies on indirect measurements of \hmol\ (e.g., \citealt{Bolatto2013}) such as virial mass, gamma-rays, dust emission, dust absorption, surrogate molecules, all of which may have uncertain calibration relationships. By measuring CO and \hmol\ together in absorption in low-optical depths transitions, and CO in emission along the same lines of sight, we can calibrate the CO-to-\hmol\ conversion factor used for emission lines 
\citep{Burgh2007,Liszt2008,Liszt2017}. By observing lines of sight of different extinction, different metallicity, and resolving the different components in the line of sight,  we can  measure X$_{\rm CO}$ as a function of  cloud depth ($A_V$) and of metallicity (which can then be applied to the low-metallicity systems in the distant universe). 
 \linebreak
It is also important to characterize the "dark CO" zones, where hydrogen is already molecular, but carbon still under C$^+$ or C$^0$. These zones typically cover A$_V$ = 0.1 to 1 \citep[e.g.,][]{Wolfire2010}, which can represent a significant amount of molecular gas.
\paragraph{Excitation of \hmol} 
Population of \hmol\ in the  J$>$2 rotational levels in the standard diffuse ISM  is not well understood: is it due to optical pumping radiative excitation \citep{Gillmon2006} or the presence of warm \hmol\ \citep{Verstraete1999,Gry2002,Falgarone2005} ?
Its understanding requires a
better characterisation of the gas through the
measure of temperature and turbulence of \hmol\ in the high-$J$ levels. 
At a resolving power higher than 10$^5$ distinctive signatures of individual components or cloud regions with different excitation could be identified. 
Some models invoke shocks
or the dissipation of turbulence in vortices to produce the J$>$2 \hmol\ as well
as CH$^+$ \citep{Godard2014}. These scenarios  imply specific
signatures in the velocity distribution of the warm \hmol\ gas, which could be
searched for in high resolution spectra. \linebreak
In several cases an  increase in velocity dispersion with rotational excitation level $J$ has been observed \citep[often at high redshift because of higher resolution,][]{Noterdaeme2007,Klimenko2015}. Towards $\zeta$OriA (observed at high resolution with IMAPS) \cite{Jenkins.Peimbert1997} have interpreted it as \hmol\ being created in a post-shock zone via the formation of H$^-$.
 \linebreak 
 Other interpretations involve energy being transferred to vibrational and rotational excitation upon \hmol\ formation. 
If a significant fraction goes into ejecting the newly formed molecules at large velocity \citep[fast \hmol\ production,][]{Barlow.Silk1976}, \cite{Jenkins.Wallerstein1999} have shown that this would produce detectable broad wings in the high-J \hmol\ profiles, provided the resolution is high enough to differentiate them from the slow \hmol.  Observing the detailed velocity dispersions enable discussing what proportions of the $4.5$\,eV energy  available  is transferred to \hmol\ excitation, to kinetic motion of the molecules, or in the form of heat for the grain. 
\paragraph{Small-scale structures in interstellar clouds:}
Knowing both the velocity and spatial structure is critical to describe the \hmol\ self shielding and the H-to-\hmol\ transition. 
The spatial structure can be studied through repeat observations of lines of sight as they drift through the foreground clouds due to the motions of the target star or the observer \citep{Boisse2005,Lauroesch2007} 
for instance through observations of runaway stars or binary stars. One needs to observe at very high signal-to-noise ratio and high resolution atoms or various molecules like \hmol, CO, and CN, CH, CH$^+$. See \cite{Welty2007} for multi-epoch optical and UV 
  observations of variable absorption 
  in C\1 fine-structure lines tracing variations in local n$_H$.
Tracking of spatial and temporal absorption variations  provide a better understanding of the nature and the properties of tiny-scale atomic structures  \citep[$\sim$10$^{1-4}$AU,][]{Heiles1997} thought to be part of a universal turbulent cascade \citep{Stanimirovic.Zweibel2018}). 
Such observations in the far-UV enable the detection of structures down to potentially much lower H\1 column densities as compared to H\1\ 21 cm absorption surveys. 
\section{Diffuse ionized gas}
{\bf Ionization structure} \,\,\, Studying the ionization structure of the warm diffuse medium involves getting detailed ionization fractions as a function of ionizing energy. This is possible through the observations of different ionization stages like [S\1, S\2, S\3, S\4, S\6], [O\1, O\6], [C\1, C\2, C\3, C\4], [Si\1, Si\2, Si\3, Si\4], [N\1, N\2, N\3, N\5] and H\1. Many of them have lines  in the UV domain, however important stages only have lines in the far-UV domain. It is therefore imperative to get access to both the UV and far-UV domains at the same high resolution. Locally the simplicity of the short sight lines toward stars in the solar vicinity ($<100$\,pc) provides a unique opportunity to study the ionization structure of individual interstellar regions, clouds, and interfaces, that are usually blended in longer sight lines  \citep{Gry.Jenkins2017}. The detection of the  weak lines that are critical for these studies requires the possibility to record UV and far-UV spectra of hot nearby stars at high signal-to-noise ratio (SNR well in excess of 100). At larger scales observing samples of post-asymptotic giant branch (PAGB) and blue horizontal-branch (BHB) stars (V$\sim$15) toward globular clusters that also contain a pulsar (whose dispersion measure yields an integrated value for n$_e$),
provides detailed ionization fractions of the WIM as a function of ionizing energy \citep{Howk.Consiglio2012}. 
%
\linebreak
\linebreak
{\bf Partly ionized gas} \,\,\,
The ionization fraction and electron density in the neutral gas is an important parameter to examine as it controls the abundance of free electrons heating the gas and providing pathways for molecular gas formation in the gas phase. It also enables us to understand the influence of supernovae and compact objects on the ISM through the propagation of ionizing cosmic rays and soft X-rays. Partly ionized gas can be studied through species with large photoionization cross-section (e.g., N\2, Ar\1) and through the population in fine-structure levels (e.g., the N\2\ multiplet at $1084$\AA, C\2*/C\2), while temperature can, for instance be determined from Mg\1/Mg\2\ or Si\2*/Si\2\ (e.g., \citealt{Jenkins.Gry.Dupin2000,Vladilo2003,Jenkins2013,Gry.Jenkins2017}). High 
SNR is necessary to detect absorption-lines arising from fine-structure levels (e.g., Si\2*, O\1**). 
%
\linebreak
{\bf High Velocity Clouds (HVCs) in the Galactic halo} \,\,\,Understanding the origins of the HVCs is difficult when neither their distances nor their metallicities are well known. Since they can cover a broad range of ionization conditions, knowledge of the ionization corrections to be applied is key to determine the metallicities. Since the gas may well not be in equilibrium a decisive way to free oneself from uncertainties is to observe all conceivable ionization states for a same atom. 
This requires the far-UV domain, especially access to N\2 at 1084 \AA\ and  O\6 at 1032 and 1038 \AA.
 Determining the covering factor of HVCs as a function of distance and z-height by observing a large sample of halo stars (B1 to B5, PAGB, BHB) with known distances (GAIA) would help determine whether the HVCs are disrupted and incorporated into the halo coronal gas as they fall or if they survive as neutral or ionized gas and reach the disk where they can fuel star formation. Only the H\1\ Lyman lines enable  measuring $N$(H\1) down to $\lesssim10^{13}$\,cm$^{-2}$ (e.g., \citealt{Lehner2006,Zech2008}). 
\section{Highly ionized gas}
The origin and nature of the collisionally-ionized gas seen notably in O\6\ absorption in the disk and halo of our Galaxy (corresponding to $T\approx3\times10^5$\,K) is still debated  \citep{Wakker2003,Savage.Lehner2006,Otte.Dixon2006,Welsh.Lallement2008} : is it formed in radiative cooling  supernovae-shocked gas, in conductive interfaces with cold gas, in turbulent mixing layers? While information on O\6\ has been limited so far to its bulk column density and relative abundance to other species, we also need information on its velocity structure and line width to disentangle the different components and to relate them to the other high-ionization species Si\3, C\4, Si\4, N\5\ and the low-ionization species, and to infer physical conditions, in particular the temperature. The electron density can also be calculated through combined observations of O\6\ in absorption and in emission \citep{Otte.Dixon2006}. 
%
\linebreak
\linebreak
{\bf Phase boundaries} \,\,\,The boundaries between the different phases can often be quite abrupt, and it is not yet clear how they trade matter and entropy, but they are believed to play a fundamental role in cooling the hot material and hence in the Galactic evolution \citep{Hopkins2013}.  
 The interaction of hot gas and cooler material, and the way processes such as thermal conduction \citep{Gnat2010} and turbulent mixing \citep{Kwak.Shelton2010} operate in the ISM in general remain poorly constrained. As does  the dynamics of a cloud propagating in a hot bubble. 
 In a survey probing the Galactic disk, \cite{Lehner2011} showed that Si IV and C IV are found in both broad and narrow components, and the high-ion column density ratios exhibit substantial variations in most lines of sight,
which implies that very different processes operate in different environments in the Galaxy. However, the confusion caused by the overlap of many different regions  over the large distances covered by surveys (e.g. for O\6, Jenkins 1978a, b; Bowen et al. 2008) has made it difficult to get
a clear picture of the nature of these interfaces.
The possibility of much simpler lines of sight is offered within the local ISM, where a single warm, diffuse cloud accounts for most of the matter within the first 50 pc \citep{Gry.Jenkins2014}, generating a single interface with the surrounding hot bubble gas.
 Its signature
 should be observed in C\4, N\5, O\6, Si\3, Si\4\  toward hot nearby stars (white dwarfs or B stars). Up to now the detection of this interface has been elusive, mostly due to the extreme weakness of the expected absorption. SNR in excess of 200 are necessary. They are theoretically easy to reach 
with nearby hot stars, but have been often limited by the bright-object limits of UV detectors and drastic neutral density filters.
\paragraph{Dynamics of shocks}
High SNR spectra obtained toward stars can reveal
 weak components at high velocities, caused by cooling layers behind shock fronts \citep{Welty2002}. Tracking velocity shifts and line widths for ions that should appear at different locations in downstream flows offers insights on how the gas cools and recombines in a time-dependent ionization scheme. The depth of absorption features in radiative shocks is expected to be larger than a few \%\ in S\3, Si\4 and C\4\ for large enough shock velocities, and should be detectable in spectra recorded at SNR in excess of a few hundreds. 
\section{Detection of very faint lines of scarce elements}
A number of scarce elements with high astrophysical interest require very high SNR to be detected because of the weakness of their lines. Let's mention: i) the light elements like $^{10}$B, $^{11}$B \citep{Lambert1998}, $^6$Li, $^7$Li \citep{Meyer1993}, or Be \citep[undetected,][]{Hebrard1997} ; ii) The r- and s- process elements like Ga, Ge, As, Se, Kr, Cd, Sn, Pb \citep{Ritchey2018}.
It would be interesting to look for localized enhancements of these elements in regions (discovered by serendipity!) where a neutron star merger occurred in a time that is more recent than a mixing time for the ISM ; iii) Isotope ratios of atomic and molecular species. For instance HCl whose line has barely been detected at 1290 \AA\ may be split in lines of H$^{35}$Cl and H$^{37}$Cl. However the interpretation of molecular isotopes may be confused by the influence of chemical
reactions. Atomic shifts   are of order a few \kms\ in some cases, so measurements should be feasible with high resolution in cold regions where velocity dispersions are low.
\section{Conclusion}
The science requirements described in the paper dictate the following instrumental specifications : 
 \begin{tabular}{ll}
 {\bf Requirement} & {\bf Justification and comments}\\
 $\lambda_{min}$ strict=1020 \AA & H I Lyman $\beta$ 1025 \AA\  ; strongest \hmol\ Lyman bands: 1030 --1155 \AA\ ; CO up to  \\ &1455 \AA\ ;O\6 1032 \AA ; N\2 1084 \AA\ ; Ar\1 1048, 1066 \AA ; O\1 1039, 1026 \AA. \\
 $\lambda_{min}$ pref = 910 \AA & H I Lyman series down to 912 \AA\ ; \hmol\ lines 912 \AA\ -- 1155 \AA\ ;  CO lines \\ & 912 -- 1455 \AA\ ; many O\1 lines of various strengths  916 -- 988 \AA \\
  $\lambda_{max}$ = 3100 \AA & Mg\2 2800 \AA\ ; Mg I 2853 \AA\ ; OH lines 3072-3082 \AA \\ 
 R  strict $>$120 000 & Resolve line profiles from cold \hmol\ with T$\simeq$100K and V$_{turb}\simeq1.2$\kms\ ; \\ &Resolve line profiles, separate thermal and turbulent contributions for warm gas: \\&Fe II with T$\simeq$6500 K and V$_{turb}\simeq$1.0\kms; separate velocity components \\& with  $\Delta$V$\simeq$3\kms;  resolve profiles from rotational levels in \hmol\ and CO bands   \\ 
 R pref $>$ 200 000 & Resolve line profiles from cold gas with V$_{turb}\leq1.0$\kms\ ; \\&resolve isotopes with $\Delta$V$\leq$1.5\kms \\
SNR $>$ 500 & Detect faint features in a reasonable amount of time. This implies achieving  \\&detectors with limited fixed-pattern noise and that can deal  with high count rates.    \\
A$_{eff}>$6 000 cm$^{-2}$ & 3 times the HST/COS value. Needed to reach A$_V>$4. Achieved with a telescope  \\& of 5 m (resp 8 m)  if the efficiency (optical  $\times$ detector) reaches 3\% (resp 1.3\%) 
\end{tabular}
\pagebreak

\bibliographystyle{aasjournal}
\bibliography{biblio}

\begin{thebibliography}{}
\expandafter\ifx\csname natexlab\endcsname\relax\def\natexlab#1{#1}\fi
\providecommand{\url}[1]{\href{#1}{#1}}

\bibitem[{{Barlow} \& {Silk}(1976)}]{Barlow.Silk1976}
{Barlow}, M.~J., \& {Silk}, J. 1976, \apj, 207, 131

\bibitem[{{Boiss{\'e}} {et~al.}(2005){Boiss{\'e}}, {Le Petit}, {Rollinde},
  {Roueff}, {Pineau des For{\^e}ts}, {Andersson}, {Gry}, \&
  {Felenbok}}]{Boisse2005}
{Boiss{\'e}}, P., {Le Petit}, F., {Rollinde}, E., {et~al.} 2005, \aap, 429, 509

\bibitem[{{Bolatto} {et~al.}(2013){Bolatto}, {Wolfire}, \&
  {Leroy}}]{Bolatto2013}
{Bolatto}, A.~D., {Wolfire}, M., \& {Leroy}, A.~K. 2013, Annual Review of
  Astronomy and Astrophysics, 51, 207

\bibitem[{{Burgh} {et~al.}(2007){Burgh}, {France}, \& {McCandliss}}]{Burgh2007}
{Burgh}, E.~B., {France}, K., \& {McCandliss}, S.~R. 2007, \apj, 658, 446

\bibitem[{{Falgarone} {et~al.}(2005){Falgarone}, {Verstraete}, {Pineau Des
  For{\^e}ts}, \& {Hily-Blant}}]{Falgarone2005}
{Falgarone}, E., {Verstraete}, L., {Pineau Des For{\^e}ts}, G., \&
  {Hily-Blant}, P. 2005, \aap, 433, 997

\bibitem[{{Ferri{\`e}re}(1998)}]{Ferriere1998}
{Ferri{\`e}re}, K. 1998, \apj, 497, 759

\bibitem[{{Gillmon} {et~al.}(2006){Gillmon}, {Shull}, {Tumlinson}, \&
  {Danforth}}]{Gillmon2006}
{Gillmon}, K., {Shull}, J.~M., {Tumlinson}, J., \& {Danforth}, C. 2006, \apj,
  636, 891

\bibitem[{{Gnat} {et~al.}(2010){Gnat}, {Sternberg}, \& {McKee}}]{Gnat2010}
{Gnat}, O., {Sternberg}, A., \& {McKee}, C.~F. 2010, \apj, 718, 1315

\bibitem[{{Godard} {et~al.}(2014){Godard}, {Falgarone}, \& {Pineau des
  For{\^e}ts}}]{Godard2014}
{Godard}, B., {Falgarone}, E., \& {Pineau des For{\^e}ts}, G. 2014, \aap, 570,
  A27

\bibitem[{{Gry} {et~al.}(2002){Gry}, {Boulanger}, {Nehm{\'e}}, {Pineau des
  For{\^e}ts}, {Habart}, \& {Falgarone}}]{Gry2002}
{Gry}, C., {Boulanger}, F., {Nehm{\'e}}, C., {et~al.} 2002, \aap, 391, 675

\bibitem[{{Gry} \& {Jenkins}(2014)}]{Gry.Jenkins2014}
{Gry}, C., \& {Jenkins}, E.~B. 2014, \aap, 567, A58

\bibitem[{{Gry} \& {Jenkins}(2017)}]{Gry.Jenkins2017}
---. 2017, \aap, 598, A31

\bibitem[{{Hebrard} {et~al.}(1997){Hebrard}, {Lemoine}, {Ferlet}, \&
  {Vidal-Madjar}}]{Hebrard1997}
{Hebrard}, G., {Lemoine}, M., {Ferlet}, R., \& {Vidal-Madjar}, A. 1997, \aap,
  324, 1145

\bibitem[{{Heiles}(1997)}]{Heiles1997}
{Heiles}, C. 1997, \apj, 481, 193

\bibitem[{{Hopkins} {et~al.}(2013){Hopkins}, {Narayanan}, \&
  {Murray}}]{Hopkins2013}
{Hopkins}, P.~F., {Narayanan}, D., \& {Murray}, N. 2013, \mnras, 432, 2647

\bibitem[{{Howk} \& {Consiglio}(2012)}]{Howk.Consiglio2012}
{Howk}, J.~C., \& {Consiglio}, S.~M. 2012, \apj, 759, 97

\bibitem[{{Jenkins}(2013)}]{Jenkins2013}
{Jenkins}, E.~B. 2013, \apj, 764, 25

\bibitem[{{Jenkins} {et~al.}(2000){Jenkins}, {Gry}, \&
  {Dupin}}]{Jenkins.Gry.Dupin2000}
{Jenkins}, E.~B., {Gry}, C., \& {Dupin}, O. 2000, \aap, 354, 253

\bibitem[{{Jenkins} \& {Peimbert}(1997)}]{Jenkins.Peimbert1997}
{Jenkins}, E.~B., \& {Peimbert}, A. 1997, \apj, 477, 265

\bibitem[{{Jenkins} \& {Wallerstein}(1999)}]{Jenkins.Wallerstein1999}
{Jenkins}, E.~B., \& {Wallerstein}, G. 1999, in Astronomical Society of the
  Pacific Conference Series, Vol. 164, Ultraviolet-Optical Space Astronomy
  Beyond HST, ed. J.~A. {Morse}, J.~M. {Shull}, \& A.~L. {Kinney}, 118

\bibitem[{{Klimenko} {et~al.}(2015){Klimenko}, {Balashev}, {Ivanchik},
  {Ledoux}, {Noterdaeme}, {Petitjean}, {Srianand }, \&
  {Varshalovich}}]{Klimenko2015}
{Klimenko}, V.~V., {Balashev}, S.~A., {Ivanchik}, A.~V., {et~al.} 2015, \mnras,
  448, 280

\bibitem[{{Kwak} \& {Shelton}(2010)}]{Kwak.Shelton2010}
{Kwak}, K., \& {Shelton}, R.~L. 2010, \apj, 719, 523

\bibitem[{{Lambert} {et~al.}(1998){Lambert}, {Sheffer}, {Federman}, {Cardelli},
  {Sofia}, \& {Knauth}}]{Lambert1998}
{Lambert}, D.~L., {Sheffer}, Y., {Federman}, S.~R., {et~al.} 1998, \apj, 494,
  614

\bibitem[{{Lauroesch}(2007)}]{Lauroesch2007}
{Lauroesch}, J.~T. 2007, in Astronomical Society of the Pacific Conference
  Series, Vol. 365, SINS - Small Ionized and Neutral Structures in the Diffuse
  Interstellar Medium, ed. M.~{Haverkorn} \& W.~M. {Goss}, 40

\bibitem[{{Le Petit} {et~al.}(2006){Le Petit}, {Nehm{\'e}}, {Le Bourlot}, \&
  {Roueff}}]{LePetit2006}
{Le Petit}, F., {Nehm{\'e}}, C., {Le Bourlot}, J., \& {Roueff}, E. 2006, The
  Astrophysical Journal Supplement Series, 164, 506

\bibitem[{{Lehner} {et~al.}(2006){Lehner}, {Savage}, {Wakker}, {Sembach}, \&
  {Tripp}}]{Lehner2006}
{Lehner}, N., {Savage}, B.~D., {Wakker}, B.~P., {Sembach}, K.~R., \& {Tripp},
  T.~M. 2006, The Astrophysical Journal Supplement Series, 164, 1

\bibitem[{{Lehner} {et~al.}(2011){Lehner}, {Zech}, {Howk}, \&
  {Savage}}]{Lehner2011}
{Lehner}, N., {Zech}, W.~F., {Howk}, J.~C., \& {Savage}, B.~D. 2011, \apj, 727,
  46

\bibitem[{{Liszt}(2008)}]{Liszt2008}
{Liszt}, H.~S. 2008, \aap, 492, 743

\bibitem[{{Liszt}(2017)}]{Liszt2017}
---. 2017, \apj, 835, 138

\bibitem[{{Meyer} {et~al.}(1993){Meyer}, {Hawkins}, \& {Wright}}]{Meyer1993}
{Meyer}, D.~M., {Hawkins}, I., \& {Wright}, E.~L. 1993, \apj, 409, L61

\bibitem[{{Nehm{\'e}} {et~al.}(2008){Nehm{\'e}}, {Le Bourlot}, {Boulanger},
  {Pineau Des For{\^e}ts}, \& {Gry}}]{Nehme2008}
{Nehm{\'e}}, C., {Le Bourlot}, J., {Boulanger}, F., {Pineau Des For{\^e}ts},
  G., \& {Gry}, C. 2008, \aap, 483, 485

\bibitem[{{Noterdaeme} {et~al.}(2007){Noterdaeme}, {Ledoux}, {Petitjean}, {Le
  Petit}, {Srianand}, \& {Smette}}]{Noterdaeme2007}
{Noterdaeme}, P., {Ledoux}, C., {Petitjean}, P., {et~al.} 2007, \aap, 474, 393

\bibitem[{{Otte} \& {Dixon}(2006)}]{Otte.Dixon2006}
{Otte}, B., \& {Dixon}, W. V.~D. 2006, \apj, 647, 312

\bibitem[{{Ritchey} {et~al.}(2018){Ritchey}, {Federman}, \&
  {Lambert}}]{Ritchey2018}
{Ritchey}, A.~M., {Federman}, S.~R., \& {Lambert}, D.~L. 2018, The
  Astrophysical Journal Supplement Series, 236, 36

\bibitem[{{Rollins} \& {Rawlings}(2012)}]{Rollins.Rawlings2012}
{Rollins}, R.~P., \& {Rawlings}, J.~M.~C. 2012, \mnras, 427, 2328

\bibitem[{{Savage} \& {Lehner}(2006)}]{Savage.Lehner2006}
{Savage}, B.~D., \& {Lehner}, N. 2006, \apjs, 162, 134

\bibitem[{{Stanimirovi{\'c}} \& {Zweibel}(2018)}]{Stanimirovic.Zweibel2018}
{Stanimirovi{\'c}}, S., \& {Zweibel}, E.~G. 2018, Annual Review of Astronomy
  and Astrophysics, 56, 489

\bibitem[{{Verstraete} {et~al.}(1999){Verstraete}, {Falgarone}, {Pineau des
  Forets}, {Flower}, \& {Puget}}]{Verstraete1999}
{Verstraete}, L., {Falgarone}, E., {Pineau des Forets}, G., {Flower}, D., \&
  {Puget}, J.~L. 1999, in ESA Special Publication, Vol. 427, The Universe as
  Seen by ISO, ed. P.~{Cox} \& M.~{Kessler}, 779

\bibitem[{{Vladilo} {et~al.}(2003){Vladilo}, {Centuri{\'o}n}, {D'Odorico}, \&
  {P{\'e}roux}}]{Vladilo2003}
{Vladilo}, G., {Centuri{\'o}n}, M., {D'Odorico}, V., \& {P{\'e}roux}, C. 2003,
  \aap, 402, 487

\bibitem[{{Wakker} {et~al.}(2003){Wakker}, {Savage}, {Sembach}, {Richter},
  {Meade}, {Jenkins}, {Shull}, {Ake}, {Blair}, {Dixon}, {Friedman}, {Green},
  {Green}, {Kruk}, {Moos}, {Murphy}, {Oegerle}, {Sahnow}, {Sonneborn},
  {Wilkinson}, \& {York}}]{Wakker2003}
{Wakker}, B.~P., {Savage}, B.~D., {Sembach}, K.~R., {et~al.} 2003, The
  Astrophysical Journal Supplement Series, 146, 1

\bibitem[{{Welsh} \& {Lallement}(2008)}]{Welsh.Lallement2008}
{Welsh}, B.~Y., \& {Lallement}, R. 2008, \aap, 490, 707

\bibitem[{{Welty}(2007)}]{Welty2007}
{Welty}, D.~E. 2007, \apj, 668, 1012

\bibitem[{{Welty} {et~al.}(2002){Welty}, {Jenkins}, {Raymond}, {Mallouris}, \&
  {York}}]{Welty2002}
{Welty}, D.~E., {Jenkins}, E.~B., {Raymond}, J.~C., {Mallouris}, C., \& {York},
  D.~G. 2002, \apj, 579, 304

\bibitem[{{Wolfire} {et~al.}(2010){Wolfire}, {Hollenbach}, \&
  {McKee}}]{Wolfire2010}
{Wolfire}, M.~G., {Hollenbach}, D., \& {McKee}, C.~F. 2010, \apj, 716, 1191

\bibitem[{{Zech} {et~al.}(2008){Zech}, {Lehner}, {Howk}, {Dixon}, \&
  {Brown}}]{Zech2008}
{Zech}, W.~F., {Lehner}, N., {Howk}, J.~C., {Dixon}, W. V.~D., \& {Brown},
  T.~M. 2008, \apj, 679, 460

\end{thebibliography}

\end{document}